\documentclass[preprint]{aastex}

\shorttitle{Enhanced Germanium in Planetary Nebulae}
\shortauthors{Sterling et al.}

\begin{document}

\title{Discovery of Enhanced Germanium Abundances in Planetary Nebulae with the \emph{Far Ultraviolet Spectroscopic Explorer}}

\author{N. C. Sterling\altaffilmark{1}, Harriet L. Dinerstein\altaffilmark{1}, and Charles W. Bowers\altaffilmark{2}}
\altaffiltext{1}{Astronomy Department, University of Texas, Austin, TX 78712-1083}
\altaffiltext{2}{Laboratory for Astronomy and Solar Physics, Code 681, NASA Goddard Space Flight Center, Greenbelt, MD 20771}

\begin{abstract}

We report the discovery of \ion{Ge}{3}~$\lambda$1088.46 in the planetary nebulae (PNe) SwSt~1, BD+30$^{\rm o}$3639, NGC~3132, and IC~4593, observed with the \emph{Far Ultraviolet Spectroscopic Explorer}\footnote{Based on observations made with the NASA-CNES-CSA \emph{Far Ultraviolet Spectroscopic Explorer}.  \emph{FUSE} is operated for NASA by Johns Hopkins University under NASA contract NAS5-3298.}.  This is the first astronomical detection of this line and the first measurement of Ge ({\it Z} = 32) in PNe.  We estimate Ge abundances using S and Fe as reference elements, for a range of assumptions about gas-phase depletions.  The results indicate that Ge, which is synthesized in the initial steps of the {\it s}-process and therefore can be self-enriched in PNe, is enhanced by factors of $\ge$ 3 -- 10. The strongest evidence for enrichment is seen for PNe with Wolf-Rayet central stars, which are likely to contain heavily processed material.

\end{abstract}

\keywords{line: identification---planetary nebulae: general---planetary nebulae: individual (BD+30$^{\rm o}$3639, SwSt~1, IC~4593, NGC~3132)---stars: AGB and post-AGB---nucleosynthesis, abundances---ultraviolet: ISM}

\section{INTRODUCTION}

Asymptotic giant branch (AGB) stars, the progenitors of planetary nebulae (PNe), are believed to be the primary synthesis site for heavy element isotopes produced by the slow neutron-capture or {\it s}-process (K\"{a}ppeler et al. 1989; K\"{a}ppeler 1999).  Since enhanced surface abundances of \emph{s}-process products are observed in evolved red giants (e.g. Smith \& Lambert 1990), it is reasonable to expect such enrichments in PNe as well.  P\'{e}quignot \& Baluteau (1994) reported optical emission lines indicating overabundances for a large number of {\it n}-capture elements in the PN NGC~7027, and Dinerstein (2001) identified two near-infrared lines commonly seen in PNe as fine-structure transitions of the light \emph{n}-capture elements Kr and Se (\emph{Z}=34, 36).  Germanium ($Z$=32) is more abundant than Kr and Se in the solar system, and, like these elements, can be synthesized in AGB stars of roughly solar metallicity (Busso et al. 1999; 2001).

In this Letter, we report the discovery of the strong resonance line \ion{Ge}{3}~$\lambda$1088.46 in four PNe observed with \emph{FUSE}, and present evidence that Ge has been self-enriched in their precursor stars.

\section{OBSERVATIONS AND DATA REDUCTION}

\emph{FUSE} (Moos et al. 2000; Sahnow et al. 2000) observations of BD+30$^{\rm o}$3639 and SwSt~1 were obtained for General Investigator (GI) programs A085 and B069 (P.I. Dinerstein) in time-tag mode using the LWRS aperture ($30\farcs 0 \times 30\farcs 0$). NGC~3132 was observed for program P133 (P.I. Bianchi), also with the LWRS in time-tag mode, while IC~4593 was observed with the HIRS aperture ($1\farcs 25 \times 20\farcs 0$) in histogram mode for GI program B032 (P.I. Gruendl). The \emph{FUSE} data and observational information may be accessed from the MAST archive at the Space Telescope Science Institute under the indicated GI programs.

The individual exposures were calibrated with CALFUSE~v2.0.5, and all exposures for a given object were cross-correlated and co-added to increase the signal-to-noise ratio S/N. We based our subsequent analysis on the detector segments with the highest S/N in the salient wavelength regions: LiF~2A for \ion{Ge}{3} $\lambda$1088.46 and the \ion{Fe}{3} lines at $\lambda\lambda$1122-1132, and LiF~1A for \ion{S}{3}~$\lambda\lambda$1012--1021.  In cases of overlapping coverage, the results were reasonably consistent in both detector segments. The spectra were rebinned by 3 pixels, increasing the S/N near 1088 \AA\ to 15--25 per rebinned sample, and broadening the sampling interval from $\sim$~7~m\AA\ to $\sim$~20~m\AA.

The continua were normalized and flattened by fitting low-order Legendre polynomials to the continuum shape and dividing them out.  We then fit the lines with single or multi-component Gaussian profiles after Sembach et al. (1993) convolved with a single Gaussian instrumental profile of FWHM 20~km~s$^{-1}$ ($R=15,000$). The normalized profiles are shown in Figure~1, with the fits for nebular features overplotted as thick solid lines.  Other lines which were modeled to constrain the fit of the nebular features are shown as dashed lines. The dotted vertical lines indicate nebular absorption velocities from \ion{Na}{1} (Dinerstein et al. 1995) or from other lines in the FUSE spectrum.

In Table~1 we summarize the values of central velocity $<v>$ and equivalent width W$_{\lambda}$ from the line fits.  The Doppler widths $b$ are poorly determined because they are small compared to the \emph{FUSE} instrumental resolution element; typical fitted values were $b\sim$10~km~s$^{-1}$. The errors are 1$\sigma$ values based on uncertainties in continuum placement and statistical errors in the fitting process; upper limits are 3$\sigma$. Wavelengths and data for atomic transitions are from Morton (1991, 2000). Data on H$_2$ transitions were obtained from S. McCandliss (private communication based on Abgrall et al.
1993 a,b; also see http://www.pha.jhu.edu/$\sim$stephan/H2ools/h1h2data/). 

\section{RESULTS}

\subsection{Identification of $\lambda$1088.46 as \ion{Ge}{3}}

We identify the feature centered at rest wavelength $\lambda$1088.46 as an intrinsically strong \ion{Ge}{3} line (oscillator strength $f$=1.84, Morton 2000). Since Ge$^{++}$ is located in the \ion{H}{2} region, it should have the same velocity as other ionized species.  SwSt~1 and IC~4593 show good agreement (Table 1), while the \ion{Ge}{3} lines appear at slightly more negative velocities in BD+30$^{\rm o}$3639 and NGC~3132. However, these discrepancies are probably not significant given the resolution of \emph{FUSE}.

Since the ionization potentials (IPs) of \ion{Ge}{2} and \ion{Ge}{3} are 15.9 eV and 34.2 eV respectively, we expect \ion{Ge}{3} to contain a large fraction of the gas-phase Ge in PNe with relatively cool central stars, where \ion{O}{2} comprises a significant fraction of O.  (In comparison, IP(\ion{O}{1}) = 13.6 eV and IP(\ion{O}{2}) = 35.1 eV.)  SwSt 1 and BD+30$^{\rm o}$3639 have cool ($T~\sim$~35,000 K), late Wolf-Rayet ([WR]) type central stars and relatively high (O$^+$)/(O$^{++}$).  IC~4593 also has a fairly cool central star (Kaler \& Jacoby 1991). The central star of NGC~3132 is a visual binary consisting of an A2~V star plus a hotter, low-luminosity companion (M\'{e}ndez 1978).  We set upper limits on $N$(Ge$^{+3}$) based on non-detections of \ion{Ge}{4} $\lambda$1229.84 ($f$=0.25) in STIS spectra of SwSt~1 (De Marco et al. 2002) and BD+30$^{\rm o}$3639, and on $N$(Ge$^{+}$) from $\lambda$1237.06 ($f$=1.23) for SwSt~1 and BD+30$^{\rm o}$3639, and $\lambda$1098.71($f$=0.55) for IC~4593 and NGC~3132.

We found no plausible alternate identifications for the feature near $\lambda$1088.46.  The nearest lines are \ion{Cl}{1}~$\lambda$1088.05, with a wavelength offset corresponding to $-$111 km~s$^{-1}$ relative to the rest wavelength of \ion{Ge}{3}, and H$_2$~P(4)~2-0 $\lambda$1088.79, at +91 km~s$^{-1}$, neither of which interfere with \ion{Ge}{3} in these objects.  This wavelength region also contains several lines from $v\ge1$ levels of H$_2$, e.g. Lyman~2-3~$R(0)$, 2-3~$R(1)$, 0-1~$R(11)$, and 9-1~$R(11)$. Of our four targets, only BD+30$^{\rm o}$3639 and possibly NGC~3132 contain observable amounts of vibrationally excited H$_2$. However, unblended lines from these levels with similar $f$-values have W$_{\lambda}$~$\le$~15~m\AA\ in BD+30$^{\rm o}$3639 (Dinerstein et al., in preparation), and SwSt~1, with the strongest $\lambda$1088.46 feature, does not show nebular H$_2$ absorption at all.  We therefore conclude that the observed features are primarily due to \ion{Ge}{3}.

We also searched the archival \emph{FUSE} spectra of 23 other PNe, but found no other detections of \ion{Ge}{3}. However, various factors could be responsible for this, such as a weak UV continuum, central star hot enough to ionize most of the Ge beyond Ge$^{++}$, or lack of sufficiently large column of ionized gas along the line of sight.  Many PNe have asymmetrical structures, and because these measurements require a bright FUV continuum there is a selection bias in favor of nebulae for which the central star is seen along a low-extinction, low column-density sight line. 

\subsection{Determination of Gas-Phase Column Densities}

The column densities listed in Table~1 were computed from the measured W$_{\lambda}$ values, assuming that the lines are on the linear part of the curve of growth. In fact, some of the lines are marginally saturated. If the line widths were purely thermal ($b$ = 1.5 km~s$^{-1}$ for Ge, 2.3 km~s$^{-1}$ for S), the line center optical depths would be $\tau_0~\sim$~5.  However, the fitted $b$ values are $\sim$ 10 km~s$^{-1}$, probably due to turbulence or shear motions, so that $\tau_0~<$ 1 except for lines from the ground levels of the more abundant ions.  In those cases we use lines from excited fine-structure levels, which are less saturated, to infer populations in the ground level.

The quantity of most interest is the Ge abundance with respect to an element which is unaffected by internal nucleosynthesis.  Whenever possible, we use S as the reference element since it is not depleted into dust (Savage \& Sembach 1996), and assume that (Ge/S) $\sim$ (Ge$^{++}$)/(S$^{+}$~+~S$^{++}$), as \ion{S}{2} and \ion{S}{3} together span a similar ionization range as \ion{Ge}{3} (15.9--34.2 eV for \ion{Ge}{3}; 10.4--23.3 eV for \ion{S}{2} and 23.3--34.8 eV for \ion{S}{3}).  We take \ion{Fe}{3} (IP range 16.2--30.7 eV) as the reference species for NGC~3132 since we did not detect \ion{S}{3} (see below).

Whereas \ion{Ge}{3} has a 4s$^2$ ground configuration and therefore only one low-lying energy level, \ion{S}{3} and \ion{Fe}{3} possess several fine-structure levels which may contain significant populations. Ideally one should measure absorption lines from each level, but this is not always feasible.  For \ion{S}{3}, lines from both excited levels could be measured in all but NGC~3132.  We used the web-based multi-level atom program of Shaw \& Dufour (1995; http://stsdas.stsci.edu/nebular) to predict the population ratios among the three fine-structure levels at the temperature and density of the ionized gas, and compared these with our derived column densities. If the measured population ratio of the excited levels agreed with the multi-level calculation, we use the latter to obtain the ground level population and total ionic column density.
  
SwSt~1 is the only object for which \ion{S}{2} and transitions from all three levels of the ground $^3$P term of \ion{S}{3} are observable. The lines from the excited levels, \ion{S}{3}* $\lambda$1015.78 and \ion{S}{3}** $\lambda$1021.33, are free of blending, but the corresponding transition from the ground level at $\lambda$1012.50 is potentially saturated and contaminated by several strong resonance ($v$ = 0) H$_2$ lines.  The value derived from this line is about 40\%\ lower than the ground-level population inferred from the excited levels, which we attribute to saturation. We used \ion{S}{2} $\lambda$1250.58, a low-$f$ line in the STIS band, to derive $N$(\ion{S}{2}). 

For BD+30$^{\rm o}$3639, \ion{S}{3} $\lambda$1012.50 is not observable due to strong H$_2$ absorptions from the nebula and foreground ISM, so $N$(\ion{S}{3}) in Table~1 is from \ion{S}{3}~$\lambda$1190.21.  However, since this line is saturated, we computed the ground-state population from the excited levels as described above, yielding a value about 2.6 times higher than measured. All of the \ion{S}{2} lines are saturated, so we adopt \ion{S}{2}/\ion{S}{3}~=~0.79 from Pwa et al. (1986) to infer $N$(\ion{S}{2}). We used a similar approach to determine ground state and total \ion{S}{3} column densities for IC~4593, and obtained $N$(\ion{S}{2}) by assuming the S ionic ratio from optical emission lines (Bohigas \& Olgu\'{i}n 1996).

Since S is not detected in NGC~3132, we were forced to use the less satisfactory species \ion{Fe}{3}. The $^5$D ground term of \ion{Fe}{3} has several fine-structure levels, although the populations of the higher levels are likely to be small due to their low statistical weights.  We marginally detect $\lambda$1124.88, the strongest ($f$=0.052) line from the first excited level, $^5$D$_3$. The surprising strength of $\lambda$1130.40 from the $^5$D$_0$ level suggests that this line may be blended or incorrectly identified.  However, not only is the ground-level $\lambda$1122.52 line saturated, it is also contaminated by \ion{C}{1} and possibly by vibrationally-excited H$_2$. Since these effects tend to offset each other and the measurements of the excited levels are uncertain, we did not apply a multi-level calculation. For these reasons as well as the uncertain depletion of gas-phase Fe (see \S 3.3), our abundance estimate of Ge in NGC~3132 is less certain than for the other PNe. 

\subsection{Correction for Dust Depletion}

The degree to which each element is depleted out of the gas and into the solid (dust) phase is an important issue for interpreting our results. As mentioned above, S is not significantly depleted in the ISM (Savage \& Sembach 1996). On the other hand, Fe is depleted by varying amounts in different components of the ISM: --1.25~dex in the warm ionized medium (WIM); as much as --2.27~dex in cold, neutral material (CNM; Savage et al. 1992).  The Ge depletion has been measured along only a handful of sight lines; a value of --0.62~dex was determined by Cardelli et al. (1991), and Welty et al. (1999) found similar values, with slightly less depletion in WIM than in CNM gas. 

Table~2 summarizes the Ge abundances we derive for each PN relative to the indicated reference ions. The two listed values correspond to the depletion patterns of the CNM and WIM, where we take the Fe depletions cited above. For Ge, we assume --0.62~dex for the CNM case, and no depletion for the WIM, providing lower limits to the Ge abundances. It seems probable that conditions in PNe are most similar to those of the WIM, or intermediate, but the values in Table~2 cover the likely range of values.

\subsection{Ge Abundances and Implications} 

The derived Ge abundances in these objects range from $\sim$ 3--30 times solar, depending on the object and assumed depletion factors.  We emphasize that the depletion of Ge is particularly uncertain, and that our values might be underestimates.  Regardless, the qualitative result of this study is that Ge significantly enriched in the observed PNe.

It is interesting that the two best detections of \ion{Ge}{3}, in SwSt~1 and BD+30$^{\rm o}$3639, involve PNe with [WR] central stars.  The surface compositions of these stars have presumably been altered by their prior evolution, since they exhibit severe H deficiencies and C enrichments.  While the detailed evolutionary path which produces a [WR] central star is not well understood, it is likely to involve late thermal pulses, deep mixing, and/or heavy mass loss (Iben et al. 1993; Herwig 2001; Bl\"{o}cker 2001).  Therefore it is plausible that nucleosynthetic products such as {\it s}-process nuclei may be enhanced to a greater degree in nebulae with [WR] central stars than in other PNe.  

\section{SUMMARY}
 
We report the discovery of \ion{Ge}{3}~$\lambda$1088.46 in four PNe observed with the \emph{Far Ultraviolet Spectroscopic Explorer}.  This line, which arises in the ionized zone, is used to estimate the abundance of Ge in the nebular gas relative to the elements S and Fe, whose abundances are not altered by the star's evolution. We find convincing evidence for elevated abundances $\ge$3--10 times solar (depending on assumed depletion factors) and therefore self-enrichment of Ge by {\it s}-process nucleosynthesis in the progenitor stars.  This result demonstrates the potential of UV absorption-line spectroscopy to shed light on the operation of the {\it s}-process in PN progenitor stars.  By determining the abundances of various heavy elements in PNe, one can set constraints on models of stellar nucleosynthesis and mixing in late stellar evolutionary stages, and directly investigate the process of chemical enrichment of the ISM by stars that evolve through the AGB and PN phases.

We are grateful to the \emph{FUSE} operations and science teams, whose exceptional efforts have made this facility available to the astronomical community.  We thank S. McCandliss for providing data on H$_2$, O. De Marco for sharing her STIS spectrum of SwSt~1, and D. Lambert and C. Sneden for helpful comments.  Financial support was provided by NASA contracts NAG5-9239, NAG5-11597, and NSF grant AST 97-31156.

\begin{deluxetable}{llccc}
\tablecolumns{5}
\tablewidth{0pc} 
\tablecaption{Line Fit Parameters}
\tablehead{
\colhead{} & \colhead{} & \colhead{$<v>$} & \colhead{$W_{\lambda}$} & \colhead{$N$} \\
\colhead{Object} & \colhead{Ion} & \colhead{(km s$^{-1}$)} & \colhead{(m\AA)} & \colhead{(cm$^{-2}$)}}
\startdata
SwSt 1..... & \ion{Ge}{3} $\lambda$1088.46 & --42.2$\pm$0.5 & 56$\pm$5 & (3.0$\pm$0.3)$\times10^{12}$\\
 & \ion{Ge}{2} $\lambda$1098.71 & \nodata & $< 9.4$ & $< 1.6\times10^{12}$ \\
 & \ion{Ge}{4} $\lambda$1189.03 & \nodata & $< 90$ & $< 1.4\times10^{13}$ \\
 & \ion{S}{3} $\lambda$1012.50 & --42.3$\pm$1.3 & 72$\pm$8 & (2.2$\pm$0.4)$\times10^{14}$\\
 & \ion{S}{3}$^*$ $\lambda$1015.78 & --38.6$\pm$0.4 & 64$\pm$7 & (4.8$\pm$0.6)$\times10^{14}$\\
 & \ion{S}{3}$^{**}$ $\lambda$1021.33 & --47.6$\pm$0.9 & 61$\pm$8 & (5.6$\pm$0.6)$\times10^{14}$ \\
 & \ion{S}{2} $\lambda$1250.58\tablenotemark{a} & --39.2$\pm$2.9 & 33$\pm$4 & (4.4$\pm$0.6)$\times10^{14}$\\
BD+30$^{\rm o}$3639... & \ion{Ge}{3} $\lambda$1088.46 & --82.3$\pm$0.6 & 35$\pm$4 & (1.8$\pm$0.3)$\times10^{12}$\\
 & \ion{Ge}{2} $\lambda$1237.06 & \nodata & $< 15$ & $< 8.9\times10^{11}$ \\
 & \ion{Ge}{4} $\lambda$1229.84 & \nodata & $< 20$ & $< 6.0\times10^{12}$ \\
 & \ion{S}{3} $\lambda$1190.21 & --66.1$\pm$1.0 & 99$\pm$7 & (2.7$\pm$0.2)$\times10^{14}$ \\
 & \ion{S}{3}$^*$ $\lambda$1015.78 & --75.1$\pm$0.5 & 47$\pm$3 & (2.4$\pm$0.4)$\times10^{14}$\\
 & \ion{S}{3}$^{**}$ $\lambda$1021.33 & --75.1$\pm$0.7 & 53$\pm$12 & (2.1$\pm$0.6)$\times10^{14}$\\
NGC 3132.... & \ion{Ge}{3} $\lambda$1088.46 & --68.1$\pm$2.1 & 30$\pm$6 & (1.6$\pm$0.3)$\times10^{12}$\\
 & \ion{Ge}{2} $\lambda$1098.71 & \nodata & $< 20$ & $< 4.4\times10^{12}$ \\
 & \ion{Fe}{3} $\lambda$1122.52 & --52.6$\pm$1.0 & 130$\pm$18 & (2.2$\pm$0.3)$\times10^{14}$\\
 & \ion{Fe}{3}$^*$ $\lambda$1124.88 & --52.8$\pm$2.0 & 9.5$\pm$1.8 & (1.6$\pm$0.3)$\times10^{13}$\\
 & \ion{Fe}{3}$^*$ $\lambda$1130.40 & --49.8$\pm$0.7 & 63$\pm$5 & (7.2$\pm$0.7)$\times10^{13}$\\
IC 4593..... & \ion{Ge}{3} $\lambda$1088.46 & --2.9$\pm$2.0 & 9.6$\pm$1.5 & (6.3$\pm$1.4)$\times10^{11}$\\
 & \ion{Ge}{2} $\lambda$1098.71 & \nodata & $< 6.2$ & $< 1.0\times10^{12}$ \\
 & \ion{S}{3}$^*$ $\lambda$1015.78 & --5.5$\pm$0.5 & 39$\pm$6 & (2.9$\pm$0.7)$\times10^{14}$\\
 & \ion{S}{3}$^{**}$ $\lambda$1021.33 & --5.5\tablenotemark{b} & 20$\pm$8 & (8.1$\pm$4.1)$\times10^{13}$\\
\enddata
\tablenotetext{a}{Measured from STIS spectra.}
\tablenotetext{b}{Due to the weakness of \ion{S}{3}$^{**}$ $\lambda$1021.33,
the central velocity was fixed to be the same as for \ion{S}{3}$^*$ $\lambda$1015.78.}
\end{deluxetable}

\clearpage

\begin{deluxetable}{llcccc}
\tablecolumns{6}
\tablewidth{0pc}
\tabletypesize{\small} 
\tablecaption{Relative Germanium Abundances} \tablehead{ \colhead{} & \colhead{Reference} & \multicolumn{2}{c}{$N$(\ion{Ge}{3})/$N(X^{+i})$} & \multicolumn{2}{c}{[Ge/X]} \\ \cline{3-4} \cline{5-6}
\colhead{Object} & \colhead{Ion} & \colhead{CNM\tablenotemark{a}} & \colhead{WIM\tablenotemark{b}} 
 & \colhead{CNM\tablenotemark{a}} & \colhead{WIM\tablenotemark{b}}}
\startdata
SwSt 1 & \ion{S}{2}, \ion{S}{3} & (6.9$\pm$2.2)$\times10^{-3}$ & (1.6$\pm$0.5)$\times10^{-3}$ & 1.65$\pm$0.15 & 0.86$\pm$0.15 \\
BD+30$^{\rm o}$3639 & \ion{S}{2}\tablenotemark{c}, \ion{S}{3} & (3.6$\pm$1.3)$\times10^{-3}$ & (8.6$\pm$3.1)$\times10^{-4}$  & 1.18$\pm$0.17 & 0.57$\pm$0.17 \\
NGC 3132 & \ion{Fe}{3} & (1.6$\pm$0.4)$\times10^{-4}$ & (4.1$\pm$1.0)$\times10^{-4}$ & 0.09$\pm$0.12 & 0.49$\pm$0.12 \\
IC 4593 & \ion{S}{2}, \ion{S}{3}\tablenotemark{d} & (3.6$\pm$1.5)$\times10^{-3}$ & (8.5$\pm$3.5)$\times10^{-4}$ & 1.20$\pm$0.19 & 0.57$\pm$0.19 \\
\enddata
\tablecomments{Columns (3) and (4) give column density ratios $N$(\ion{Ge}{3})/$N$(X$^{+i})$ where 
X is the indicated reference ion (S or Fe); values in square brackets in Columns (5) and (6) are
in logarithmic units with respect to the meteoritic values of Grevesse \& Noels (1993).
Cited errors are propagated from the values in Table 1 and corrections made 
for depletion and unobserved levels using Shaw \& Dufour's (1995) web-based 5-level program.}
\tablenotetext{a}{Column densities of \ion{Ge}{3} and \ion{Fe}{3} adjusted for depletions 
in the cold interstellar clouds toward $\zeta$ Oph (Savage et al. 1992; Savage \& Sembach 1996).  
Sulfur is assumed to be undepleted.}
\tablenotetext{b}{Column density of \ion{Fe}{3} adjusted for depletion seen  
  in the warm diffuse cloud toward $\zeta$ Oph 
  (Savage et al. 1992; Savage \& Sembach 1996). Ge and S are assumed to be undepleted.}
\tablenotetext{c}{Assumes (S$^{+}$/S$^{++}$)
   from Pwa et al. (1986).}
\tablenotetext{d}{Assumes (S$^{+}$/S$^{++}$) from Bohigas \& Olgu\'{i}n (1996).}
\end{deluxetable}

\clearpage

\figcaption[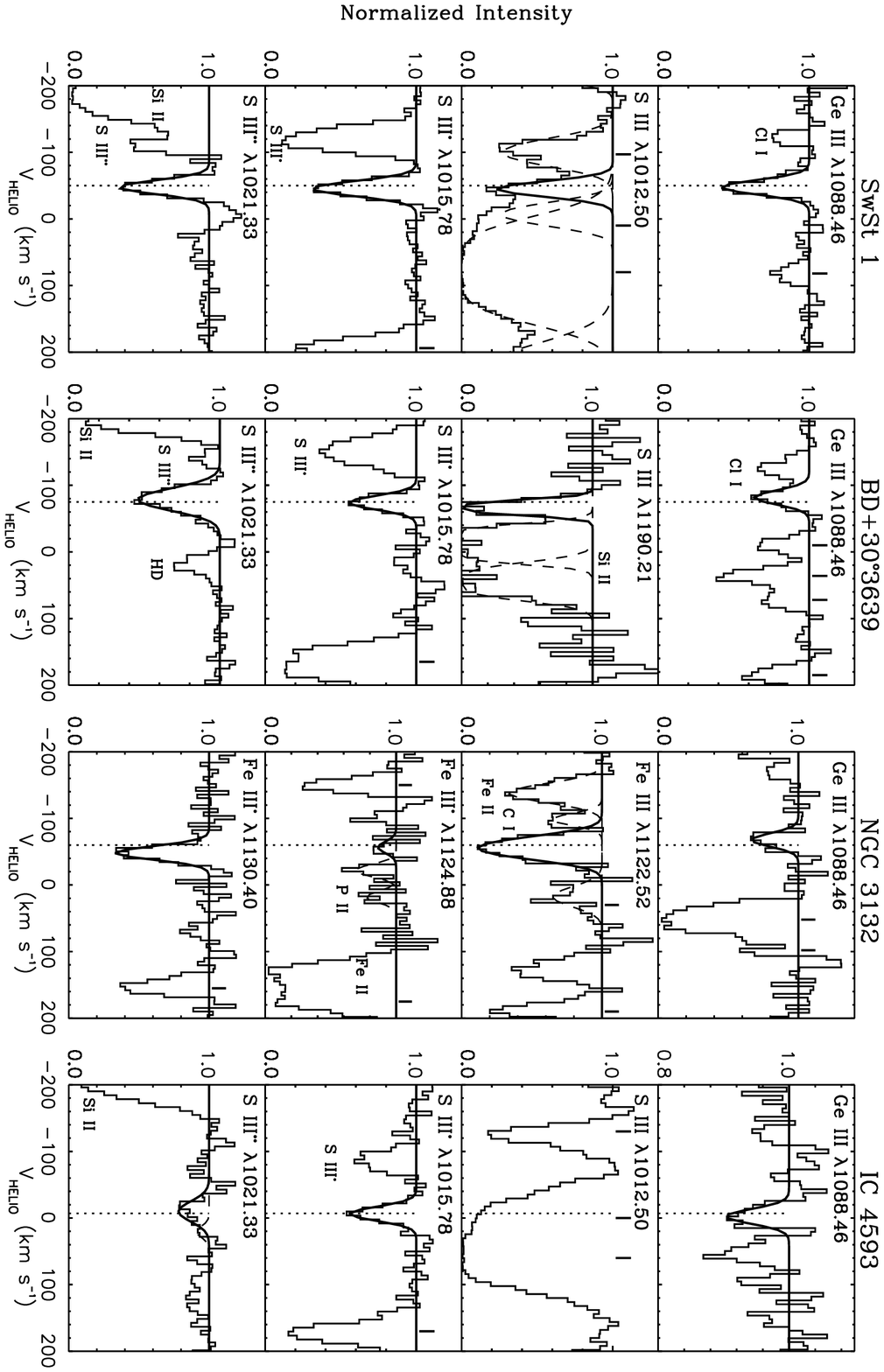]{Continuum normalized absorption spectra are plotted against heliocentric velocity for the indicated species.  Profiles for each object are arranged by column, with reference lines below \ion{Ge}{3} $\lambda$1088.46.  The \ion{S}{3}~$\lambda$1190.21 profile is from STIS data, while all other features shown were obtained by \emph{FUSE}.  Note that the vertical scale is enlarged for IC~4593, which has the weakest \ion{Ge}{3} line.  Other ionic lines are indicated by species name and H$_2$ lines by vertical tick marks.  The heavy solid lines are modeled Gaussian fits to nebular absorptions; other lines modeled in the fitting process are shown as dashed lines.  The vertical dotted lines indicate the nebular velocities measured from Na I and other species.  We were unable to fit \ion{S}{3}~$\lambda$1012.50 in IC~4593 due to blending with H$_2$.}

\clearpage

\begin{figure}
\plotone{fig1.ps}
\end{figure}

\end{document}